\documentclass{article}
\usepackage{arxiv}
\usepackage{amsmath}
\usepackage{amsfonts}
\usepackage{amssymb}
\usepackage{algorithm}
\usepackage{algorithmic}
\usepackage{graphicx}
\usepackage{balance}
\usepackage{url}

\def\SAVE#1{}
\newtheorem{theorem}{Theorem}

\newtheorem{problem}{Problem}
\newtheorem{proposition}[theorem]{Proposition}

\newtheorem{example}{Example}

\begin{document}

\title{Query Scheduling in the Presence of Complex User Profiles}

\author{Haggai Roitman\thanks{Work was done while the author was still a PhD student in the Technion - Israel Institute of Technology.}  \\
IBM Research\\
Haifa University Campus, Mount Carmel\\
Haifa 31905, Israel\\
haggai@il.ibm.com \and Avigdor Gal  \\
Technion - Israel Institute of Technology\\
Haifa 32000, Israel\\
avigal@ie.technion.ac.il \and Louiqa Raschid \\
University of Maryland\\
College Park, MD 20742\\
louiqa@umiacs.umd.edu\\
}

\maketitle

\begin{abstract}
Advances in Web technology enable personalization proxies that assist users
in satisfying their complex information monitoring and aggregation needs
through the repeated querying of multiple volatile data sources.
Such proxies face a scalability challenge when trying to maximize the number of
clients served while at the same time fully satisfying clients' complex
user profiles. In this work we use
an abstraction of complex execution intervals (CEIs) constructed over simple
execution intervals (EIs) represents user profiles and use existing offline approximation as a baseline for maximizing completeness of
capturing CEIs.
We present three heuristic solutions for the online problem of
query scheduling to satisfy complex user profiles. The first only considers
properties of individual EIs while the other two exploit properties of
all EIs in the CEI.
We use an extensive set of experiments on real traces and synthetic data to show that
heuristics that exploit knowledge of the CEIs dominate across multiple parameter settings.
\end{abstract}

\section{Introduction}

Advances in Web technology now
enable the creation of personalization proxies that assist users in satisfying
their complex information monitoring and aggregation needs. A proxy
actively decides when it needs to query information streams using pull-based
technology to satisfy clients' complex user profiles. Example platforms
include personalization portals and news aggregation applications (\emph{e.g.,}
MyYahoo!\footnote{\url{https://my.yahoo.com/}}, Feedly\footnote{\url{https://feedly.com}}), which provide a single point of
access, services for continuously refreshing profiles, and tools for
integration via a \emph{mashup}\footnote{See http://www.programmableweb.com/
for mashup examples.} of data extracted from multiple heterogeneous data
sources. Proxies are required to query multiple streams of events in a timely
manner to satisfy both the characteristics of servers, \emph{e.g.,} intensity
of updates, and the complex user profiles of clients.
This results in a scalability challenge when the proxy tries to satisfy
millions of clients.

Push-based solutions to satisfy complex user needs exist, e.g., WebSocket, WebRTC, HTTP/2 Push, etc.
Yet, data collection with
push-based solutions requires a high cost \cite{CherniackBBCCXZ03}, especially
if such push-based technology is not natural to the Web environment.
Pull-based solutions have considered only simple monitoring solutions
(\emph{e.g.}, \cite{PANDEY2004}) that cannot satisfy complex user needs.
Lately, such complex requirements have been also introduced in the context of
Web and Social streams search, discovery, and analysis (\emph{e.g.},
\cite{BansalK07,paul2016social,Roitman2008MDC,Sakaki:2010,LaFleur:2015,woodall2017}) which require the collection of data from
multiple (possibly interrelated) pull-only sources (e.g., RSS feeds).

Following~\cite{ROITMAN2008b}
, complex user profiles are represented in this work as a set of \emph{complex execution
interval}s (CEIs). A CEI is an extension of a simple
execution interval (EI) \cite{ROITMAN2008}
which defines periods of time
during which the proxy has to probe the corresponding resources to satisfy the
profile. Querying multiple data sources is represented by combining individual
EIs to construct CEIs, possibly over a set of resources. Each EI in a CEI
should be monitored once for the CEI to be satisfied (or \textquotedblleft
captured"). A proxy schedules queries to corresponding resources in these
intervals to satisfy the profile.

\begin{figure}[tbh]
\center\includegraphics[width=3.35in]{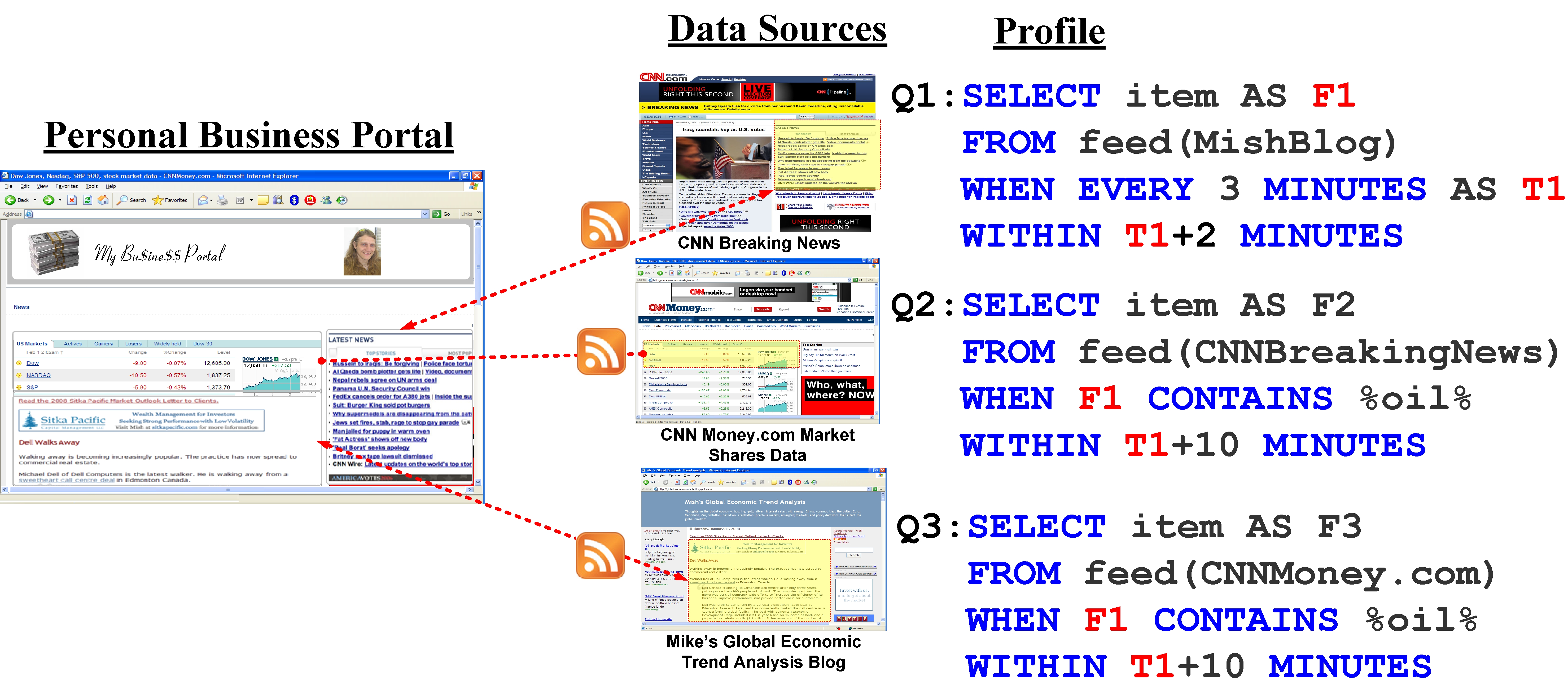}\caption{Motivating
example: personal business portal.}%
\label{Fig:ex}%
\end{figure}

We illustrate a personalized portal in Figure \ref{Fig:ex}. A business analyst
identifies data sources that fit her needs, \emph{e.g.}, the \emph{CNN
Breaking News}\footnote{\url{http://edition.cnn.com/}} website, \emph{CNN
Money.com},\footnote{\url{http://money.cnn.com/}} and \emph{Mish's Global Economic
Trend Analysis}\footnote{http://globaleconomicanalysis.blogspot.com/} blog.
The data sources can be specified as URLs of online Web feeds (\emph{e.g.,}
RSS, Atom) as shown by the RSS symbols in Figure \ref{Fig:ex}. Alternatively,
\emph{Web Scraping}\footnote{\url{http://en.wikipedia.org/wiki/Web\_scraping}.}
technology allows her to delineate content that is to be extracted. Most Web
feeds are available via pull-only access protocols, \emph{e.g.}, via
\texttt{HTTP GET} requests. However, some feeds may be pushed to the user,
with the appropriate registration, using proprietary technology (\emph{e.g.,}
Google Alerts service\footnote{\url{www.google.com/alerts}}).

The analyst constructs a wide perspective by integrating data from multiple
business and market data sources. She is interested in querying \emph{CNN
Breaking News} and \emph{CNN Money.com} when a periodic querying of
\emph{Mish's Global Economic Trend Analysis} blog detects that a new post in
the blog contains the word \emph{\%oil\%}. This will be translated into the
three queries in Figure \ref{Fig:ex}. The analyst is willing to accept a delay
of up to two minutes in probing \texttt{MishBlog} and a delay of 10 minutes
for the other two feeds.

According to \cite{HONG2005}, 55\% of Web feeds are updated hourly. Further,
about 80\% of the feeds have an average size smaller than 10 KB, suggesting
that items are promptly removed from the feeds. These statistics on refresh
frequency and volatility illustrate the challenges faced in satisfying
millions of complex profiles.

Our research goal is
to support a new generation of solutions to address complex user profiles in (primarily)
pull-only settings.

Previously, \cite{ROITMAN2008b,ROITMAN2009}
 has introduced the problem of
capturing CEIs and showed that an offline solution is of high polynomial
complexity.
An approximate \textbf{offline} solution was presented in~\cite{ROITMAN2009}
as a baseline
for maximizing completeness. In this paper, we present three \textbf{online} heuristics that
can scale to capture millions of CEIs. S-EDF is a simple extension of the
well-known Early Deadline First policy and only considers individual EIs while
making a schedule. MRSF considers the number of EIs in a CEI that have not
been scheduled and M-EDF considers the deadlines of all the EIs in the CEI
that have not been scheduled; thus, both exploit the properties of CEIs. Using
an extensive empirical analysis of real and synthetic data, we demonstrate the
dominance of the heuristics that exploit the properties of CEIs.

The paper is organized as follows:
In Section \ref{sec:model and problem definition} we summarize the model of
\cite{ROITMAN2008a,ROITMAN2009}
and define the problem. Section
\ref{sec:solution} presents the heuristic solution. We present experiments in
Section \ref{sec:experiments} and in Section \ref{sec:related} we describe
related work. Section \ref{sec:conclusions} concludes with future work.

\section{Model and Problem Definition}

\label{sec:model and problem definition}

We summarize the model of complex user profiles presented in \cite{ROITMAN2008b,ROITMAN2009},
for completeness sake. \emph{Clients} query
\emph{Servers} through \emph{proxies}. A server manages resources and can be
\emph{queried} (pull-based) by the proxy on behalf of its clients.
We discuss the three building blocks of our model, namely \emph{client
profiles}, \emph{execution intervals}, and \emph{schedules}, and then define
the problem.

\subsection{Profiles and complex execution intervals}

\SAVE{
The complex information needs of a client are specified as a profile stored
at the proxy. A client profile is a set of queries, extended with
declarative monitoring guidelines. As an example, consider Figure \ref{Fig:ex}, in which a user profile is defined as a set of three queries. The
\texttt{when} clause determines the triggering event (\emph{e.g.}, \texttt{EVERY 3 MINUTES} or \texttt{WHEN F1 CONTAINS \%}oil\texttt{\%}, with the
obvious interpretation). The \texttt{within} clause determines the slack the
proxy is given in reporting an event (\emph{e.g.}, \texttt{T1+2 MINUTES}).
}

A client's complex user profile is translated into a set of resources and
\emph{complex execution intervals} (CEI). An execution interval (EI) \cite{ROITMAN2008}
defines periods of time during which the resource must be
probed. A profile combines individual EIs, possibly over a set of resources,
to construct a CEI. Each EI in a CEI should be monitored once for the CEI to
be satisfied (or \textquotedblleft captured"). Details of a query language
to express complex user profiles is presented in \cite{ROITMAN2008b}.
We use
pseudo continuous queries in our examples. We expect that a proxy will
provide tools, \emph{e.g.}, Web scraping, to provide intuitive interfaces to
clients.

Formally, let $\mathcal{R}=\{r_{1},r_{2},...,r_{n}\}$ be a set of $n$
resources and let $\mathcal{T}=\left( T_{1},T_{2},...,T_{K}\right) $ be an
epoch with $K$ chronons.\footnote{%
A chronon is an indivisible unit of time.} We assume the proxy manages a set
of client profiles $\mathcal{P}=\{p_{1},p_{2},\ldots ,p_{m}\}$. A client
profile $p=\left\{ \eta |\eta =\langle I_{1},I_{2},\ldots ,I_{t}\rangle
\right\} $ is a collection of CEIs. A CEI $\eta $ contains several EIs,
where each EI $I$ is associated with a resource $r\in \mathcal{R}$ and $I$
contains a start and finish chronon $I=[T_{s},T_{f}];T_{s},T_{f}\in \mathcal{%
T};T_{s}\leq {T_{f}}$. An interesting class of profiles, we denote by $%
\mathcal{P}^{[1]}$, are profiles for which any EI $I$ of any CEI has a width
of \emph{exactly one chronon}. This class serves later in our analysis.
\begin{figure}[tbh]
\center\includegraphics[width=3.0in]{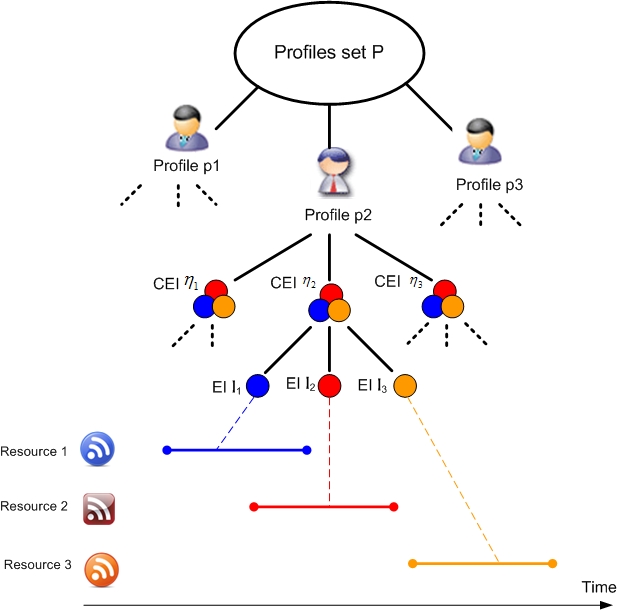}
\caption{A hierarchical illustration of the model's main concepts.}
\label{fig:concepts}
\end{figure}

Profiles, CEIs, and EIs construct a \emph{hierarchy} as seen in Figure \ref%
{fig:concepts}.
Two CEIs within the same profile, or two EIs within the same CEI, are \emph{%
siblings}. To model \emph{profile complexity}, we denote by $rank(p)$ the
maximal number of execution intervals in any CEI $\eta \in {p}$
$=\max_{\eta \in {p}}\left\{ \left\vert \eta \right\vert \right\} $, where $%
\left\vert \eta \right\vert $ is the number of execution intervals in $\eta $%
. The definition is easily extended to a set of profiles $\mathcal{P}$ as
follows: $rank(\mathcal{P})=\max_{p\in \mathcal{P}}\left\{ rank(p)\right\} $.

The beginning of an interval is determined by an update event at a resource
or a temporal event (\emph{e.g.}, every three minutes). In the case that the
server will \emph{push} the update event, or for a temporal event, the
beginning of the interval is deterministic. A proxy may also need to predict
an update event using an update model and stochastic modeling \cite{GAL2000b}.
The window (length) of the interval is determined with respect to the
stream of update events, (\emph{e.g.}, update = overwrite), or as a temporal
event (\emph{e.g.}, within five minutes of the beginning of the interval).
For example, a profile for Web scraping over Web feeds requires that
published items be collected before the server overwrites them; this is a
stochastic event.

\begin{figure}[tbh]
\center
\includegraphics[width=3.5in]{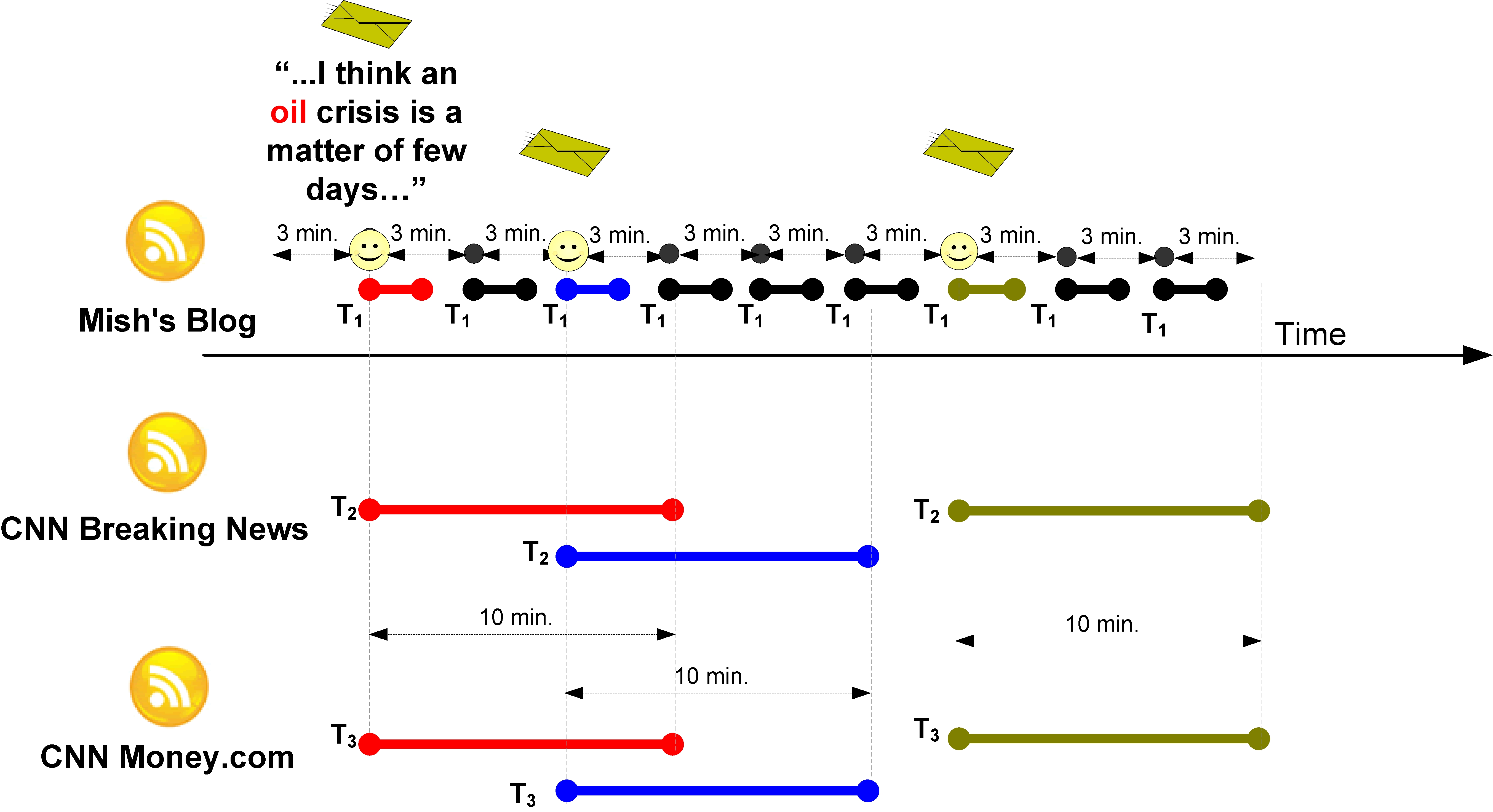}\newline
\caption{EIs for the profile of the case study example}
\label{fig:exp3}
\end{figure}

The CEIs of our motivating example (see Figure \ref{Fig:ex}) are illustrated
in Figure \ref{fig:exp3}. Probing the \texttt{MishBlog} feed every 3
minutes, with a possible slack of 2 minutes, is represented by the first set
of EIs, labeled \texttt{T1}. We note that by registering to a Pub/Sub
system, the proxy may be informed of updates to Mish's blog. However, the
proxy still has to query to get the updated blog. For posts on Mish's blog
that contain the keyword \texttt{oil}, EIs will be scheduled to probe the
other two resources, labeled \texttt{T2} and \texttt{T3}, respectively. In
this example, some CEIs will only have a rank of 1 (when updates on Mish's
blog do not include \texttt{oil}) while others will have a rank of 3.

EIs of the same or different profiles may overlap in time; two cases of
overlap are interesting. When EIs of different resources overlap (\emph{inter-resource overlap}), then they are all candidates for being
simultaneously queried by the proxy. This can lead to \emph{congestion} when
the available budget is low. When EIs associated with an \emph{identical}
resource overlap (\emph{intra-resource overlap}), there is the potential to
exploit this overlap to build an efficient query schedule. While delaying a
query on a resource could lead to more efficient schedules, a delay may also
result in failing to capture an EI, \emph{e.g.}, there is congestion-based
competition in a future chronon.
The special case of \emph{no intra-resource overlap} is of theoretical
interest since it allows us to present some bounds \cite{ROITMAN2009}.

\subsection{Schedules}

\label{schedules and metrics} A data delivery schedule $S=\{s_{i,j}%
\}_{i=1,...,n;j=1,...,K}$ ($n$ resources and $K$ chronons) assigns $%
s_{i,j}=1 $ if resource $r_{i}\in \mathcal{R}$ should be queried by the
proxy at chronon $T_{j}\in \mathcal{T}$, else $s_{i,j}=0$. We denote by $%
\mathbb{S}$ the set of all possible schedules.

The expression $\mathbb{I}(I,S)$ indicates whether a schedule $S$
successfully captures (\emph{i.e.}, some resource $r_{i}$ is queried during)
the EI $I$. Given a profile $p$, a CEI $\eta \in {p}$, and an EI $I\in \eta $
that refers to resource $r_{i}\in \mathcal{R}$, we have the following:
\[
\mathbb{I}(I,S)=\left\{
\begin{array}{cc}
1, & \exists T_{j}\in {I}:s_{i,j}=1 \\
0, & otherwise%
\end{array}%
\right.
\]

We extend $\mathbb{I}(I,S)$ to describe capturing a CEI as follows: Given a
profile $p$ and a CEI $\eta \in {p}$, we say that $\eta $ is \emph{captured}
by schedule $S\in \mathbb{S}$ if $\mathbb{I}(\eta ,S)=\prod_{I\in \eta }%
\mathbb{I}(I,S)=1$.

\subsection{Problem statement}

\label{sec:problem} We assume that the proxy has a limited amount of
resources that can be consumed for querying. In this paper we consider a
constraint similar to the one used in prior works of Web Monitoring \cite%
{PANDEY2004} and Web Crawlers \cite{WOLF2002}, where at each chronon $%
T_{j}\in \mathcal{T}$ the proxy can query up to $C_{j}$ resources. This
constraint is represented by a budget vector $\vec{C}=\left(
C_{1},C_{2},\ldots ,C_{K}\right) $.

Given a set of client profiles $\mathcal{P}=\{p_{1},p_{2},...,p_{m}\}$, the
proxy objective is to \emph{maximize gained completeness}, \emph{i.e.}, the
number of CEIs from $\mathcal{P}$ that are captured given the budget $\vec{C}
$. A CEI is successfully captured once \emph{all} of its execution intervals
are captured. Every CEI $\eta \in {p}$ that is successfully captured by the
proxy schedule (indicated by $\mathbb{I}(\eta ,S)=1$) increases the gained
completeness.

Given a schedule $S\in \mathbb{S}$, the \emph{gained completeness} (denoted
\texttt{GC} in short) from monitoring $\mathcal{P}$ during $\mathcal{T}$
according to $S$ is calculated as follows (where $\left\vert p\right\vert $
denotes the number of CEIs in profile $p$):
\begin{equation}
\mathtt{GC}(\mathcal{P},\mathcal{T},S)=\frac{\sum_{p\in \mathcal{P}%
}\sum_{\eta \in {p}}\mathbb{I}(\eta ,S)}{\sum_{p\in {\mathcal{P}}}\left\vert
p\right\vert }  \label{eq:gain compliteness}
\end{equation}

Formally, the problem of query scheduling in the presence of complex user
profiles is defined by the following constrained optimization problem.
\begin{problem}\label{maximization problem}
Given a set of profiles $\mathcal{P}$ and an epoch $\mathcal{T}$:
\[
 \begin{array}{l}
\emph{maximize}\ \texttt{GC}(\mathcal{P},\mathcal{T},S) \\
\emph{s.t.}\ \sum_{i=1}^{n}s_{i,j}\leq {C_{j}},\ \forall {j=1,2,\ldots ,K}%
\end{array}
\]
\end{problem}

Previously, \cite{ROITMAN2009}
presented two offline solutions to Problem \ref{maximization problem}. In an offline setting, the proxy is provided with \textbf{all} CEIs in $\mathcal{P}$ for $K$ chronons \textbf{in advance} and
has to determine the schedule $S$ of probing resources in $\mathcal{R}$. The
offline solutions are of high polynomial complexity~\cite{ROITMAN2009}. Yet, they provide a
baseline of optimal performance.

\section{Query Scheduling Policies}

\label{sec:solution} We now present three heuristic solutions to the problem
of query scheduling in the presence of complex user profiles. CEIs are not
known a priori and proxy decision making is done online. In our example, the
EIs for querying the \texttt{MishBlog} feed can be determined in advance,
but query scheduling for the other two streams, (\emph{CNN Breaking News}
and \emph{CNN Money.com}) depends on the contents of the first feed. At
every chronon $T_{j}$, the proxy may receive a set of new CEIs. The proxy
then has to decide which resources in $\mathcal{R}$ to probe, while
considering the set of all candidate CEIs, including those submitted prior
to $T_{j}$, which have not been completely captured yet, and the new\ set of
CEIs. We denote the set of all candidate CEIs at chronon $T_{j}$ as $\emph{%
cands}(\eta )$ and the union bag of all their EIs (termed \emph{candidate EIs%
}) as $\emph{cands}(I)=\biguplus_{\eta _{q}\in \emph{cands}(\eta )}\eta _{q}$%
. The bag notation ($\biguplus $) is used due to intra-resource overlaps.

\subsection{Policies}

\label{sec:online policies}

To determine which candidate EIs in $\emph{cands}(I)$ to choose the proxy
uses \emph{policies}. At chronon $T_{j}$, a policy $\Phi $ considers $\emph{%
cands}(I)$ and the budget $C_{j}$, and returns up to $C_{j}$ EIs to probe.
Such policies can be efficiently implemented.

Each of the policies we propose can be executed in either a \emph{%
non-preemptive} or \emph{preemptive} manner. Non-preemptive policies do not
allow new candidate CEIs to be scheduled for monitoring at chronon $T_{j}$
if previously probed CEIs need to be probed at $T_{j}$. Therefore, a
non-preemptive policy $\Phi$ first selects $I\in \emph{cands}(I)$ that
belongs to previously probed CEIs. Then, if there is any budget left, $\Phi$
selects EIs from the newly introduced CEIs. It is worth noting that even the
non-preemptive policies do not guarantee a successful capture of a CEI that
has been probed at least once. If the number of previously probed CEIs
exceeds the bandwidth budget, some of these CEIs may be dropped.

Policies can be classified according to
the amount of information about candidate CEIs they use. We propose a three
level classification, as follows.

\noindent {\textbf{Individual EI level}}: An individual EI level policy
utilizes only the local properties of a \emph{single EI} without considering
the parent CEI or sibling EIs.\newline
\noindent As a representative of this level we suggest the \emph{Single
Interval Early Deadline First} (or \texttt{S-EDF} in short) policy, $\Phi _{%
\mathtt{S}\text{\texttt{-}}\mathtt{EDF}}$. This policy is modeled on the
well known EDF policy \cite{Liu73}, prefering EIs that have the earliest
deadline. Given an execution interval $I$ and a chronon $T$, the deadline is
calculated (in terms of number of remaining chronons) as follows:
\[
\mathtt{S}\text{\texttt{-}}\mathtt{EDF}(I,T)=I.T_{f}-T+1
\]%
Proposition \ref{prop:SEDFOptimality}
holds
for this policy.
\begin{proposition}
\label{prop:SEDFOptimality}
Given $\mathcal{P}$ without intra-resource overlap, and
$rank(\mathcal{P})=1$, the policy $\Phi_{\mathtt{S}\text{\texttt{-}}\mathtt{EDF}}$ is
optimal.
\end{proposition}

WIC \cite{PANDEY2004}, a well-known monitoring solution for the Web, can
also be classified as an individual EI level policy. We note that WIC was
designed to address a different optimization goal compared to our Problem %
\ref{maximization problem}). WIC's solution balances completeness with
timeliness, providing a bound of 2-competitiveness for its optimization
goal. WIC defines a utility for each EI and picks those EIs with the maximum
accumulated utility for probing in each chronon. For our experiments in
Section \ref{sec:experiments} we implemented WIC (details provided later)
and compared the performance of WIC with the online policies from this
section.

\noindent {\textbf{Rank level}}: A rank level policy bases its decision on
profile complexity by considering the rank of the parent CEI. As a
representative of this level we suggest the \emph{Minimal Residual Stub First%
} (\texttt{MRSF}) policy, $\Phi _{\mathtt{MRSF}}$. This policy prefers EIs
that belong to parent CEIs with a minimal number of EIs left to be captured.
The intuition behind this policy is that a CEI with less EIs remaining to
probe has a higher probability of success. Formally, given an EI $I$, $I\in
\eta $ and $\eta \in {p}$, then the \texttt{MRSF} value is calculated as
follows:%
\[
\mathtt{MRSF}(I)=rank(p)-\sum_{I^{\prime }\in \eta }\mathbb{I}(I^{\prime
},S)
\]%
where $I^{\prime }$ iterates over all EIs in $\eta $. The following
proposition provides a bound for the performance of this policy.
\begin{proposition}
Given $\mathcal{P}$ without intra-resource overlap and
$rank(\mathcal{P})=k$, the $\Phi_{\texttt{MRSF}}$ policy is
$l$-competitive, where:
\[
l=\max_{\eta\in\mathcal{P}}(\sum_{I\in\eta}\left|I\right|).
\]
\end{proposition}

\noindent {\textbf{Multi-EI level}}: A multi-EI level policy utilizes the
properties of all EIs of a parent CEI (including sibling EIs). As a
representative, we suggest the \emph{Multi Interval EDF} (\texttt{M-EDF})
policy, $\Phi _{\mathtt{M}\text{\texttt{-}}\mathtt{EDF}}$, which prefers
execution intervals that have the minimal \texttt{M-EDF} value, calculated
as follows (for $I\in \eta $):
\[
\mathtt{M}\text{\texttt{-}}\mathtt{EDF}(I,T)=\sum_{I^{\prime }\in \eta
}\left( \mathtt{S}\text{\texttt{-}}\mathtt{EDF}\left( I^{\prime },T\right)
\cdot \left[ 1-\mathbb{I}\left( I^{\prime },S\right) \right] \right)
\]

The \texttt{M-EDF} value combines the EDF values of execution interval $I$
and its siblings that were not captured by chronon $T$. For each $I^{\prime }
$ (iterating over all EIs in the CEI) if the EI is not yet active (chronon $%
T<I^{\prime }.T_{s}$), then the EDF value is calculated with $T=0$, taking
into account the full length of the EI.
The intuition is to favor CEIs with less total remaining chronons so as to
increase their probability of capture, leading to a gain in completeness.

The following proposition discusses the class $\mathcal{P}^{[1]}$ of
profiles. Recall that we denote by $\mathcal{P}^{[1]}$ a set of profiles for
which any EI $I$ of any CEI has a width of \emph{exactly one chronon}.

\begin{proposition}\label{prop. MRSF equals MS-EDF}
For problem instances with $\mathcal{P}^{[1]}$ profiles the
\texttt{M-EDF} policy is equivalent to the \texttt{MRSF} policy.
\end{proposition}

\begin{figure}[tbh]
\center
\includegraphics[width=3.35in]{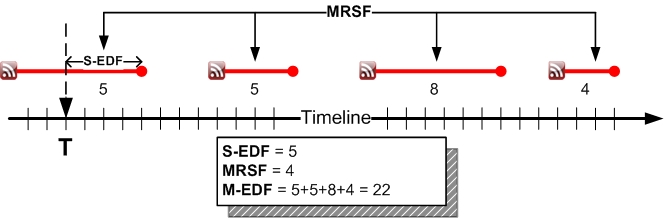}\newline
\caption{Policies illustration}
\label{policies}
\end{figure}

\begin{example}\label{example policies}
Figure \ref{policies} illustrates the value each policy assigns to a candidate
CEI with four execution intervals. At
chronon T, \texttt{S-EDF} counts the number of remaining chronons until the end
of the execution interval ($5$). \texttt{MRSF} counts the number of
remaining EIs in the CEI ($4$). Finally, \texttt{M-EDF}
accumulates the number of chronons of all remaining EIs ($22$). The values each policy assigns are given at
the bottom of the figure.
\end{example}

\begin{figure}[tbh]
\center
\includegraphics[width=2.7in]{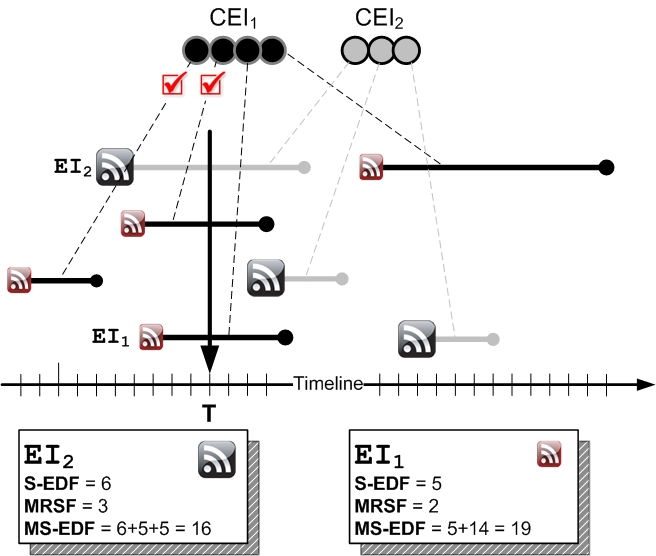}\newline
\caption{Competing CEIs illustration}
\label{fig:competingPolicies}
\end{figure}

\begin{example}\label{example policies1}
Figure \ref{fig:competingPolicies} illustrates the decision making
of each of the policies, when faced with multiple complex execution
intervals and a bandwidth constraint. Here, we have two candidates,
the first ($CEI_1$) with four required EIs (marked in black) and
the second ($CEI_2$) with three EIs (marked in gray).
Dashed lines connect a parent complex execution interval to its
children. Assume the current time is $T$ and only one EI can be
selected ($C_T=1$). At $T$ the first two EIs of $CEI_1$ were
successfuly captured (see the $\surd$ signs near the top of the
figure) yet we allow preemption. A decision needs to be taken
whether to probe $EI_1$ or $EI_2$.

\texttt{S-EDF} counts the number of remaining chronons until the end
of the current EI, yielding $5$ for $EI_1$ and $6$ for $EI_2$.
Therefore, the \texttt{S-EDF} policy will stick with $CEI_1$.

\texttt{MRSF} counts the number of remaining EIs. With $2$ remaining
EIs for $CEI_1$ and $3$ remaining EIs for $CEI_2$, \texttt{MRSF} will stick
with $CEI_1$, choosing $EI_1$.

Finally, \texttt{M-EDF} accumulates the number of chronons in all
the remaining EIs. We have $19$ chronons for $CEI_1$ and $16$ for
$CEI_2$. Therefore, $CEI_1$ will be preempted, since \texttt{M-EDF}
favors $EI_2$.
\end{example}

\section{Experiments}

\label{sec:experiments}

\subsection{Datasets and Experiment Setup}

\label{sec:dataSet}

\subsubsection{Datasets}

We used two data streams of update events in our experiments. The first is a
real-world trace of
732 auctions for Intel, IBM, and Dell laptop computers\footnote{\url{http://git.annonymous.com}.}.
To obtain the traces, we used an RSS feed retrieved for a search query on
a popular e-commerce website
to obtain timestamps of
updates and we reconstructed the real stream. Each auction lasted 3 days with a total of 11,150 bids in the datasets. We also used a synthetic data
stream that was generated using a Poisson based update model; the parameter
$\lambda$ controls the update intensity of each resource.

\subsubsection{Profiles and CEIs}

\label{sec:profile generation}

We used a profile template to specify complex user needs and to generate
multiple profile instances. \textquotedblleft$\mathtt{AuctionWatch(}%
k\mathtt{)}$" is a sample template that monitors the prices of $k$ auctions
and notifies the user after \emph{a new bid is posted in all $k$ auctions}.
The start of an EI to monitor an auction will be prediced from an update model
constructed from the traces. The length of each EI can be specified as
\texttt{overwrite} or \texttt{window(W)}. The \texttt{overwrite} requests
\emph{every new bid to be delivered before the next update occurs and
overwrites the last published bid}.

The \texttt{window(W)} requests \emph{\ every new bid to be delivered within a
window of \texttt{W} chronons from the time the bid was posted}.

We generated up to $m$ profile instances from a template using a 2-stage
process and 2 Zipf distributions. Recall that $rank(\mathcal{P})=k$
corresponds to $k$ EIs in a CEI. We determine the rank of each profile
instance according to a Zipf$(\beta,k)$ distribution, where $k$ is specified
in the template. $\beta=0$ implies a random selection or a uniform
distribution $U[1,k]$, while a positive $\beta$ value produces more profiles
whose CEIs include less ($<$ $k$) EIs. This stage models
a variance of the complexity of a profile (termed \emph{intra-user
preferences}).

In a second step, given some profile of rank $k$, we use a Zipf$(\alpha,n)$
distribution to select a set of resources. $\alpha=0$ implies a random
selection or a uniform distribution $U[1,n]$, while positive $\alpha$ value
implies a preference towards \textquotedblleft popular" resources. This stage
models \emph{inter-user preferences} and imitates the way popular resources
are chosen by users. For Web Feeds the value of $\alpha$ was estimated to be
$1.37$ \cite{HONG2005}.

\subsubsection{Experimental Setup}

We implemented a simulation-based environment to test the different solutions.
Given a profile template and an update event stream, we generate $m$ profile
instances and their CEIs.
The proxy receives as input at each chronon the set of CEIs that overlap in
that chronon.
We repeated each execution
$10$ times and recorded the average performances. Table
\ref{controlled parameters} summarizes the control parameters. We report on
two metrics, namely \emph{completeness} (as given in Eq.
\ref{eq:gain compliteness}) and \emph{runtime costs}, the execution time
normalized over the total number of EIs that must be captured.

For comparison
we implemented WIC \cite{PANDEY2004}.
We used a straightforward implementation of the algorithm presented in
\cite{PANDEY2004}, setting \texttt{urgency} to be uniform (that is, for each
resource $r_{i}$ and chronon $T$ $\mathtt{urgency}_{i}(T)=1$), \texttt{life}
to be either \texttt{overwrite} or \\ \texttt{time-window-append(W)}
\cite{PANDEY2004} (which corresponds to our \texttt{window(W)}), and $p_{ij}$
set to $1$ if resource $r_{i}$ has an update at chronon $T_{j}$, otherwise we
set $p_{ij}=0$.
While we note that WIC does not handle complex CEIs, for certain parameter
settings, \emph{i.e.}, with $rank(\mathcal{P})=1$, $\mathtt{W}=0$, and with no
intra-resource overlap, both WIC and \texttt{S-EDF} provide an optimal
schedule \cite{PANDEY2004}. Therefore both \texttt{S-EDF} and WIC are used as
a baseline of simple EI policies for comparison. In addition, we implemented
an offline approximation for the problem, as introduced in \cite{ROITMAN2009}
, which serves as another baseline for evaluating the performance of the
proposed policies.

The various policies and the experimental setup were implemented in Java, using JDK version 7. The JVM was initiated with a
heap-memory size of $1.00GB$.

\begin{table}[ptb]
\center
{
\begin{tabular}
[c]{|c|c|c|c|}\hline
\textbf{Parameter} & \textbf{Name} & \textbf{Range} & \textbf{Baseline
value}\\\hline\hline
\texttt{W} (chronons) & Max. EI length & $[0,20]$ & $10$\\\hline
$n$ & Number of Resources & $[100,1000]$ & $1000$\\\hline
$m$ & Number of Profiles & $[100,2500]$ & $100$\\\hline
$K$ & Number of Chronons & $1000$ & $1000$\\\hline
$C$ & Budget limitation & $[1,5]$ & $1$\\\hline
$\lambda$ & Avg. updates intensity & $[10,50]$ & $20$\\\hline
$rank(\mathcal{P})$ & Max. profile rank & $[1,5]$ & $3$\\\hline
$\alpha$ & Inter preferences & $[0,2]$ & $0$\\\hline
$\beta$ & Intra preferences & $[0,2]$ & $0$\\\hline
$\Phi$ & Policy & All & All\\\hline\hline
\end{tabular}
}\caption{Controlled Parameters}%
\label{controlled parameters}%
\end{table}

\subsection{Comparative policy performance}

\label{sec:proof of concept} \begin{figure}[ptb]
\center
\includegraphics[width=3.35in]{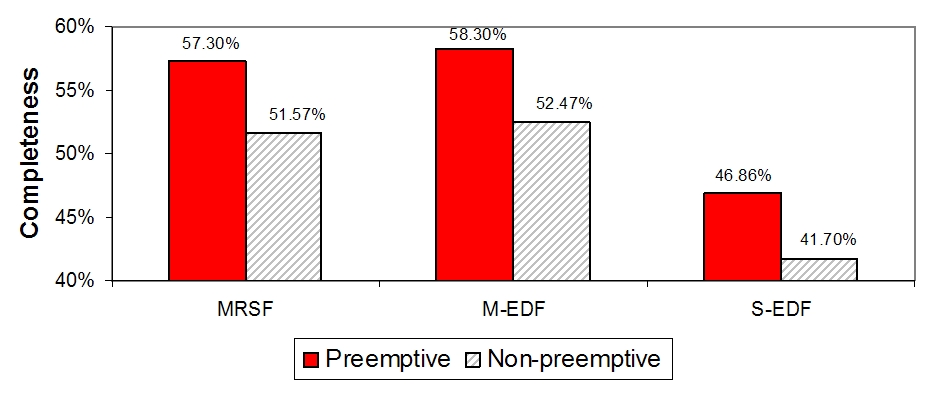}\caption{Online policies with and
without preemption}%
\label{fig:poc}%
\end{figure}

We have two versions of each of the online policies, namely with and without
preemption. To compare them, we used the real-world trace and the profile
template $\mathtt{AuctionWatch}$$(3)$ (each profile will monitor up to $3$
auctions). We also used a window of \texttt{W}=20, \emph{i.e.}, the new bid
must be reported within 20 chronons. Finally, we used a budget of $C=2$ probes
per chronon. We compared the gained completeness of each online policy using
various parameter settings, with and without preemption. Figure \ref{fig:poc}
reports on the results with $400$ auction resources, $1590$ CEIs and $3599$
simple EIs.

Typically \texttt{MRSF} and \texttt{M-EDF} perform better then \texttt{S-EDF}
over a range of experiment settings. This improved performance can be
attributed to the use of more knowledge about the rank of the profile and the
number of EIs to be captured. Our complete set of experiments with
\texttt{S-EDF} indicates that it performs better without preemption for $C=1$,
while for $C>1$, the preemptive version is better. The other two policies,
\texttt{MRSF} and \texttt{M-EDF}, perform better with preemption in all but a
few cases. These results were consistent for most of the parameter settings
that were tested, with a difference of up to $20\%$ in completeness between
the preemptive and non-preemptive versions of each online policy. We label the
policies \textquotedblleft\texttt{(P)}" to denote the preemptive version and
\textquotedblleft\texttt{(NP)}" for non-preemptive.

\subsection{Policies vs. the Baseline}

We compared the performance of the proposed policies and WIC, with the offline
approximation solution, for different parameter settings. We used the real
trace and the $\mathtt{AuctionWatch}$$(k)$ profile template with
$\mathtt{W}=0$, \emph{i.e.}, the client requires an \emph{immediate} probing
of each EI as soon as a bid is posted.

We first examine $\mathcal{P}^{[1]}$ profiles. To generate these profiles and
CEIs, we use $C=1$ and avoid intra-resource overlap by ensuring that each EI
of a CEI refers to a distinct resource. For this parameter setting, the
offline approximation guarantees a $2k$-approximation~\cite{ROITMAN2009} (for $rank(\mathcal{P}%
^{[1]})=k$). 
\ref{prop. MRSF equals MS-EDF}, both \texttt{MRSF(P)} and \texttt{M-EDF(P)}
policies perform the same and we only report on \texttt{MRSF(P)}.

\begin{figure}[ptbh]
\center
\includegraphics[width=3.0in]{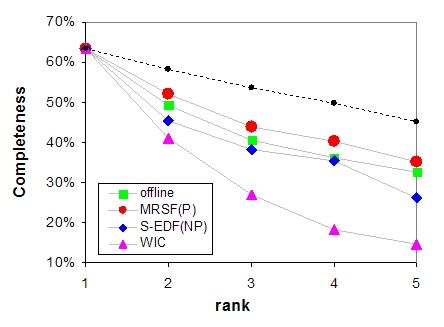}\newline\caption{Online policies
compared with the baseline}%
\label{fig:off_vs_on_rank}%
\end{figure}

Figure \ref{fig:off_vs_on_rank} reports on the results, where the
$rank(\mathcal{P})$ varies from 1 to 5. For this experiment, if
$rank(\mathcal{P})=k$, then all CEIs that were generated for that problem
instance have exactly $k$ EIs. Figure \ref{fig:off_vs_on_rank} and experiments
with other settings show that both \texttt{M-EDF(P)} and \texttt{MRSF(P)}
policies typically dominate the offline approximation, the \texttt{S-EDF}
policy (preemptive or not), and WIC. The \texttt{MRSF(P)} outperforms the
offline approximation by up to $10\%$. The \texttt{S-EDF} policy does not
dominate the offline approximation yet it dominates WIC, as is the offline
approximation. We observe that complex profiles with higher rank force the
proxy to devote more probes to each CEI and completeness decreases for all policies.

We can compute an upper bound on the optimal completeness for
$rank(\mathcal{P})=1$ by using \texttt{S-EDF} optimal completeness for the
same problem instance. We can also determine a worst case upper bound on the
optimal completeness for $rank(\mathcal{P})>1$. To do that, for every
$rank(\mathcal{P})$ level, we measure the completeness in terms of single EIs
that are captured (\emph{i.e.}, assuming that $rank(\mathcal{P})=1$); This
measure provides a worst case upper bound on the optimal performance. This
bound is illustrated by the dashed line in Figure \ref{fig:off_vs_on_rank}.
Therefore, we observe that the completeness achieved by the complex policies
is actually very high for every profile complexity level (from optimal
completeness for $rank(\mathcal{P})=1$ to at least $77\%$ of the optimal
completeness for $rank(\mathcal{P})=5$). To conclude, the complex policies
dominate both the offline approximation and the best possible simple policies
available in the literature, and provide good performance guarantees.

\subsection{Runtime scalability}

\label{sec:scalability} \label{sec:scale}

\begin{figure}[ptb]
\center
\includegraphics[width=2.75in]{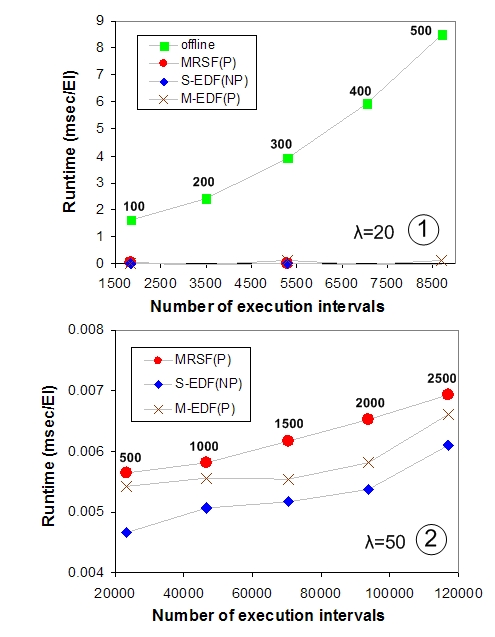}\caption{Scalability
analysis}%
\label{fig:scale uniform utility}%
\end{figure}

We next performed scalability analysis and measured the runtime of the offline
approximation and the online policies aggregated over a run with $K=1000$
chronons. The results are given in Figure \ref{fig:scale uniform utility}. The
runtime is given per a single EI processing unit. We ran the experiments with
increasing number of profiles, generating as a result an increasing number of
execution intervals. The number of profiles and the update intensity that were
used are given on top of the curves, and on the lower right side of the graphs
respectively. First we compared the offline approximation and online policies
with small workloads ($\lambda=20$ and 100-500 profiles; see Figure
\ref{fig:scale uniform utility}(1)). We can see that the offline approximation
performs much worse then the online policies. We further continued to
investigate the scalability of the online policies and increased the workload
using 2.5 times higher updates intensity and increased the number of profiles
up to 2500 (see Figure \ref{fig:scale uniform utility}(2)). We can see that
there is still a linear trend in the policies runtime behavior, suggesting
that for real problem instances the online policies are scalable and more
robust compared to the offline approximation.

\subsection{Workload analysis}

\label{sec:on vs. off} \begin{figure}[tbh]
\center
\includegraphics[width=2.7in]{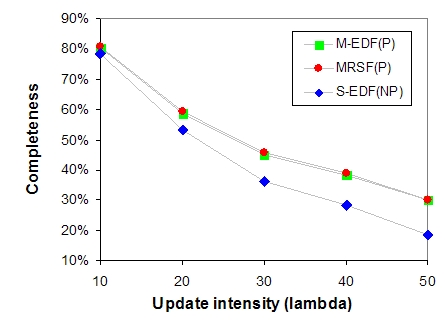}\caption{Workload analysis (update
intensity)}%
\label{fig:workload:lambda}%
\end{figure}

We next studied the effect of different workload settings on the gained
completeness of policies. For this purpose we controlled two parameter
settings, namely the average updates intensity per resource (given by
$\lambda$), and number of profiles ($m$) (where the other parameters are those
described in Table \ref{controlled parameters}). It is worth noting that for
this analysis we used a strict budget allocation of $C=1$. In Section
\ref{sec:budget} we show how a more lenient budget improves performance.

Figure \ref{fig:workload:lambda}
contains the results of this analysis for increasing update intensity.
In general, we observe that both the \texttt{MRSF(P)} and \texttt{M-EDF(P)}
policies are much better then \texttt{S-EDF(NP)} policy for all workload
parameter settings that were used. We also observe that the \texttt{MRSF(P)}
performs slightly better then \texttt{M-EDF(P)}.
Finally, as the average update intensity increases, there are more updates to
each resource, and thus, each profile requires to capture more CEIs, resulting
in decreased gained completeness.
The effect of number of profiles the proxy has to consider, on the policies
performance (not presented here due to space considerations) is similar.

\subsection{The impact of user preferences}

\label{sec:prefs} \begin{figure}[tbh]
\center
\includegraphics[width=2.7in]{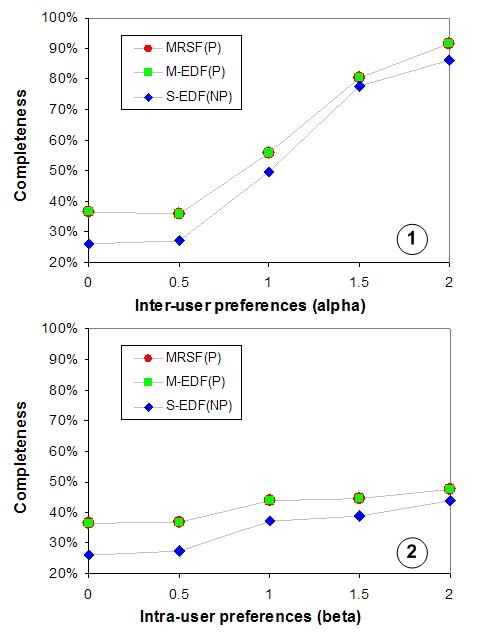}\caption{Impact of user
preferences}%
\label{fig:prefs}%
\end{figure}We now report on the impact of user preferences on performance.
For this analysis, we have used various settings of $\alpha$ (inter-user
preferences), and $\beta$ (intra-user preference), where the other parameters
were set for the baseline values of Table \ref{controlled parameters}. See
Section \ref{sec:profile generation} for the definition of these two parameters.

Figure \ref{fig:prefs}(1) shows the impact of inter-user preferences. As
$\alpha$ increases, there is less random selection of resources in each
profile, with more execution intervals coming from popular resources. The
online policies gain more completeness due to more opportunities to capture
intra-resource overlapping execution intervals of popular resources.
Furthermore, we observe that the \texttt{S-EDF(NP)} policy is dominated by the
two other policies.

Figure \ref{fig:prefs}(2) shows the impact of intra-user preferences. As
$\beta$ increases, users prefer less complex profiles. The impact, as observed
in Figure \ref{fig:off_vs_on_rank} as well, is an increase in performance. We
further observe that both \texttt{MRSF(P)} and \texttt{M-EDF(P)} policies
still outperform the \texttt{S-EDF(NP)} policy, with a slight variation
between the first two. This may be attributed to the fact that there are still
CEIs that require to probe more then one execution interval, where
\texttt{S-EDF(NP)} was shown in Figure \ref{fig:off_vs_on_rank} to be dominated.

\subsection{Effect of budgetary limitations}

\label{sec:budget} \begin{figure}[ptb]
\center
\includegraphics[width=2.7in]{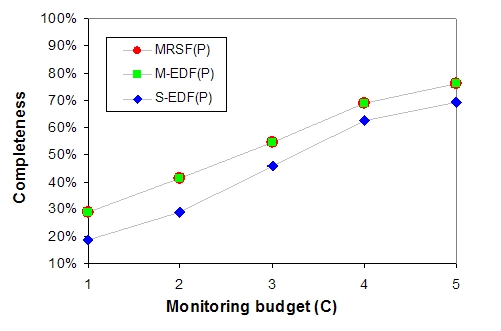}\caption{Budget limitations}%
\label{fig:budget}%
\end{figure}We studied the effect of budgetary limitations on the different
policies. So far we have used a \textbf{strict budgetary allocation }of $C=1$.
We now show the impact of additional budget on performance. The results are
given in Figure \ref{fig:budget}. We observe that as the proxy budget
increases, allowing it to probe more resources per chronon, a remarkable
increase in performance is achieved. In particular, \texttt{MRSF(P)} policy
utilizes the budget much better then the \texttt{S-EDF(P)} policy. We conclude
that the aggregated view of \texttt{MRSF(P)} and \texttt{M-EDF(P)} policies
utilizes the budget better.

\section{Related Work}

\label{sec:related}

Many contemporary applications, such as Web crawlers and monitors \cite{PANDEY2004,WOLF2002}, news feeds aggregators \cite{AltinelBCKLMMNSS07,Gruhl06,LaFleur:2015} and social streams monitors~\cite{agarwal2014time,Mathioudakis:2010,osborne2014real,paul2016social,Sakaki:2010,woodall2017},
may require nowadays complex access patterns to multiple Web sources.
We can classify these applications based on their information gathering
strategy.

With push based systems, data is pushed to the system and the research focus
is mainly on aspects of efficient data processing, where load shedding
techniques \cite{Chi05} can be applied to determine the portions of the data
that must be processed. Examples are publish-subscribe (pub-sub) (\emph{e.g.,%
} \cite{DemersGHRW06,DiaoRF04}), stream processing (\emph{e.g.,} \cite{ArasuBBDIRW03,CherniackBBCCXZ03}), and complex event
processing (CEP) (\emph{e.g.,} \cite{WuDR06,DemersGPRSW07}). Pub-sub systems
such as the ONYX system \cite{DiaoRF04} allow the registration of complex
requirements at servers and focus mainly on the trade-off between data
processing efficiency and the expressiveness of the queries that can be
processed by the system. Stream processing systems are also push-based in
nature and focus mainly on smart filtering and load shedding techniques.
Complex event processing (CEP) systems (e.g.,~\cite{DemersGPRSW07})
assume the pushing of a stream of raw events and focus
mainly on efficient complex event identification.

In our work, we assume a pull based solution; there is a need to collect the
data, \emph{e.g.,} via periodic queries. In the presence of stream data that
is available via pull-only access (\emph{e.g.,} Web feeds), complex queries
must cross multiple streams. Example systems include query processing in
sensor networks (\emph{e.g.,} \cite{DeshpandeGMHH04,Diao07}), continuous
queries (CQ) and Web monitoring (\emph{e.g.,} \cite{LiuPT00WebCQ,Garg04,PANDEY2004,ECKSTEIN2007}), Grid and Web
services query processing (\emph{e.g.,} \cite{ZhangARS05,Srivastava06}), and
mashups of Web sources (\emph{e.g.,} \cite{AltinelBCKLMMNSS07}). Current
pull based monitoring is unable to handle complex data needs over multiple
data sources. For example, current works in CQ and Web monitoring such as
WIC \cite{PANDEY2004} handle only simple single resource monitoring tasks
that are assumed to be independent of each other.

\section{Conclusions and Future work}

\label{sec:conclusions} In this work we presented efficient query scheduling
online policies for the satisfaction of complex data needs in pull based
environments that involve volatile data. Using intensive experiments we
analyzed the performance of these policies under different settings and
showed that even under restrictive budget constraints they can perform well.
We further showed that utilizing additional profile structures can assist to
improve performance significantly.

The\textbf{\ }problem presented in this work can be extended in several
interesting directions. First, we do not consider the varying costs of
probing resources. These variations may be due to computational costs, \emph{%
e.g.}, extracting a stock price may be cheaper than searching for a keyword
in a blog, the bandwidth needed to download results of varying size,
monetary charges at the servers, \emph{etc}. In this case, queries will not
consume the same budget. A second extension is to consider more general
profile satisfaction constraints given as client profile utilities. Such
utilities can help to construct better prioritized policies.

\balance
\bibliographystyle{plain}

\end{document}